\documentclass{statsoc}

\usepackage[dvips]{graphicx}
\usepackage{subfigure}
\usepackage{psfrag}
\usepackage{amsmath}
\usepackage{amssymb}
\usepackage{enumerate}
\usepackage{natbib}

\newcommand{\eref}[1]{(\ref{#1})}

\newcommand{\E}{\textrm{E}}

\title[Experimental Designs in Switching Measurements]{Experimental Designs for Binary Data in Switching Measurements on Superconducting Josephson Junctions}
\author[J. Karvanen {\it et al.}]{Juha Karvanen\thanks{Juha Karvanen,  Department of Health Promotion and Chronic Disease Prevention, National Public Health Institute, Mannerheimintie 166, 00300 Helsinki, Finland, juha.karvanen@ktl.fi}}
\email{juha.karvanen@ktl.fi}
\address{
National Public Health Institute, Helsinki, Finland}
\author[J. Karvanen {\it et al.}]{Juha J. Vartiainen, Andrey Timofeev and Jukka Pekola}
\address{
Helsinki University of Technology, Espoo, Finland}
\begin{document}

\begin{abstract}
We study the optimal design of switching measurements of small Josephson junction circuits which operate in the macroscopic quantum tunnelling regime.  Starting from the D-optimality criterion we derive the optimal design for the estimation of the unknown parameters of the underlying Gumbel type distribution. As a practical method for the measurements, we propose a sequential design that combines heuristic search for initial estimates and maximum likelihood estimation. The presented design has immediate applications in the area of superconducting electronics implying faster data acquisition. The presented experimental results confirm the usefulness of the method.
\end{abstract}

\section{Introduction}
Randomness plays a central role in quantum physics. On the fundamental level,
randomness is not caused only by unknown covariates or measurement error
but is understood to be an intrinsic property of Nature itself~\citep{Ballentine}. From the statistical point of view, this leads to fascinating problems where the distributional assumptions arise
 directly from the fundamental laws of physics.

The Josephson junctions (JJ) are important non-linear components of superconducting electronics. A typical junction involves two superconducting electrodes separated by an insulator gap~\citep{Josephson}. In practice, miniature JJ circuits are manufactured by lithographic means and operated at cryogenic temperatures, i.e., below 4.2~K. The strong dependence of the physical parameters of JJ circuits as function of changes in environmental variables, for instance, temperature, electric noise, and magnetic field makes the JJ circuits to have several applications as ultra-sensitive sensors~\citep{pekola}. Moreover, certain JJ circuits are promising candidates for realization of quantum computation~\citep{schon}. For a thorough review of JJ circuits see, for instance, ~\citep{single}.

\begin{figure}
\begin{center}
 \includegraphics[width=0.9\columnwidth]{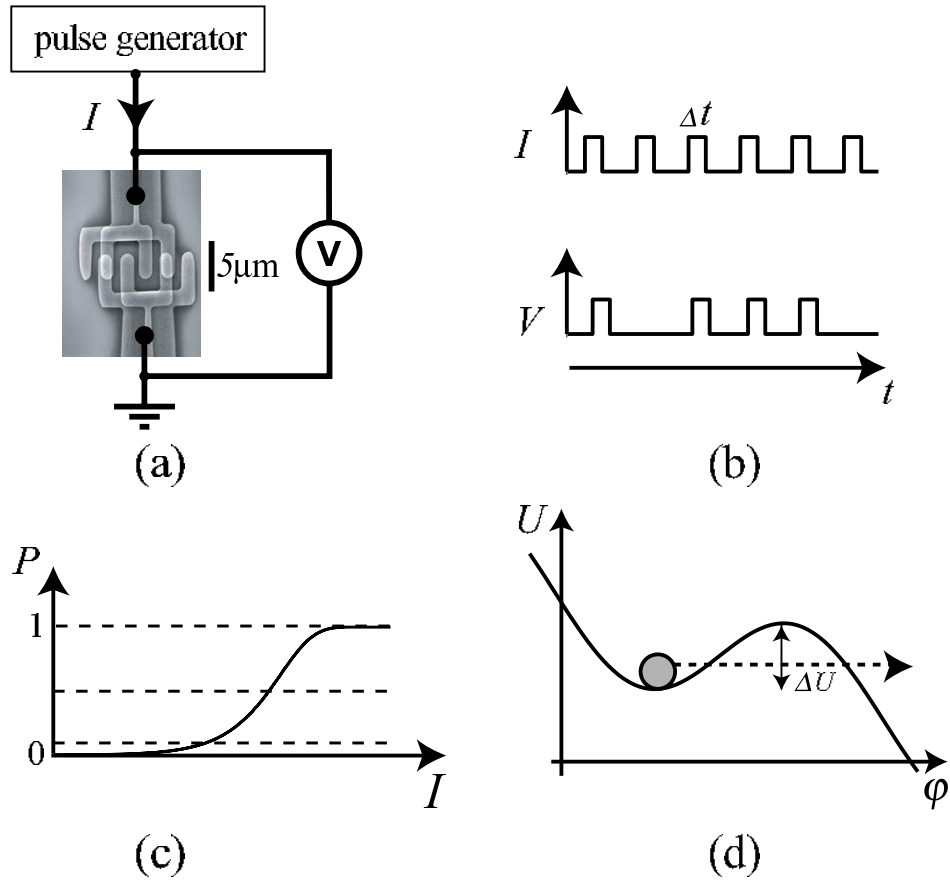}
\caption{\label{fig:expsetup} Illustration of a switching measurement.
(a) Scanning electron micrograph of a typical two-junction
JJ sample.
(b) Sketch of applied current ($I$) pulses and resulting ($V$) voltage pulses.
(c) The associated switching probability as a function of the height of the current pulses. (d) Quantum mechanical process leading to switching: The phase ``particle'' is initially localized in a local minimum of the potential. When the current pulse is applied, the phase particle tunnels through the potential barrier and escapes downhill.}
\end{center}
\end{figure}

An experiment called switching measurement, see Fig.~\ref{fig:expsetup}, is a common way to probe the properties of a JJ circuit sample. In the experiment, sequences of current pulses are applied to the sample, while the voltage over the structure is monitored. The superconducting electrodes of the sample are described by a complex valued quantum mechanical wave function and the physical state of the JJ is described by the phase difference over the junction. The dynamics of the phase difference are analogous to those of a massive particle in a tilted cosine potential, see Fig.~\ref{fig:expsetup}(d).
A sufficiently high current pulse causes the phase difference to escape from the local minimum of the potential to the direction of the steepest descent.
In the experiment the observed voltage pulse indicates that switching has occurred. The appearance of a voltage pulse to a single applied current pulse, being governed by the laws of quantum mechanics,
is purely random. Below, we call observed voltage pulse by $Y=1$, otherwise $Y=0$. The analysis of the statistics of this random variable with respect to current pulse heights yields information about the physical parameters of a JJ circuit.

We study the switching dynamics of a single JJ. Quantum mechanical arguments studied in more detail in Section~\ref{sec:physics} suggest that
the switching rate is of the form $\gamma=e^{ax+b}$, where
 $x=x(I)$ is a monotonic function depending on the height $I$ of the applied current pulse and
$a$ and $b$ are the parameters depending on the geometry and environment of the junction.
Moreover, we note that $x(I)$ must be independent of parameters $a$ and $b$.
Assuming constant switching rate during the pulse we get the probability of measuring a
voltage pulse at an applied current pulse
\begin{equation} \label{eq:yprob1}
P(Y=1)=1-e^{-\exp(ax+b)}.
\end{equation}
For the simplicity we consider the case $x(I)=I$ in this paper. In statistics,
model~\eref{eq:yprob1} is often called complementary log-log regression.

Assume that a sequence of $n$ current pulses of height $I_i, \; i=1,2,\ldots,n$ is applied to the sample. The obtained data consist of covariate values $x_{i}, \; i=1,2,\ldots,n$ and corresponding
binary responses $y_{i}$. Our aim is to choose the covariates such that the parameters $a$ and $b$ can be estimated as well as possible from the measured data. In statistical terms this is an optimal design
problem~\citep{Atkinson:optimumbook,Pukelsheim:optimalbook,Liski:optimalbook,Berger:appliedoptimal}. The literature on optimal designs is concentrated on linear models but work has been done also with nonlinear models. Several authors
\citep{Kalish:optimaldesign,Abdelbasit:experimentaldesign,Minkin:optimal,Sitter:optimal,Mathew:optimal} have considered optimal designs for binary data in the case of logistic regression
\begin{equation} \label{eq:logistic}
P(Y=1)=\frac{1}{1+e^{-(ax+b)}},
\end{equation}
where $a$ and $b$ are unknown parameters and $x$ is a covariate. Compared to the logistic regression the complementary log-log regression~\eref{eq:yprob1} is less frequently studied or applied. Complementary log-log regression is often used as an alternative model for logistic regression~\eref{eq:logistic} if the symmetry assumption does not fit to the data, but in our application the complementary log-log regression arises directly from the physical properties of JJs. The key reference considering optimal design for complementary log-log regression is the work by \citet{Ford:canonicaloptimal}. They provide the general framework for the D and c optimal designs and the solution for complementary log-log regression is obtained as a special case of their general results.

In this paper we propose a sequential two-point design for estimating the unknown parameters of a JJ circuit. The design is derived using D-optimality criterion, which equals to minimizing the generalized variance of the parameters. Based on the data available, the design yields the optimal current pulse heights for the next measurement in real time. We are ultimately interested in minimizing the time needed for an experiment. The proposed sequential design also takes into account the special requirements of switching measurements, e.g. the time needed for the reconfiguration of the signal generator.

The rest of the paper is organized as follows: In Section~\ref{sec:physics}, a more detailed description of physical background of the switching measurements is given. In Section~\ref{sec:doptimal}, the optimal design for the complementary log-log regression defined in equation~\eref{eq:yprob1} is derived. In Section~\ref{sec:sequential}, we propose a sequential design for switching measurements. In Section~\ref{sec:real}, the sequential design is applied to switching measurements of a JJ circuit. Results from the real data are compared with simulation results.
Finally, Section~\ref{sec:conclusion} concludes the paper.

\section{Physical background} \label{sec:physics}
\begin{figure}
\begin{center}
 \includegraphics[width=0.9\columnwidth]{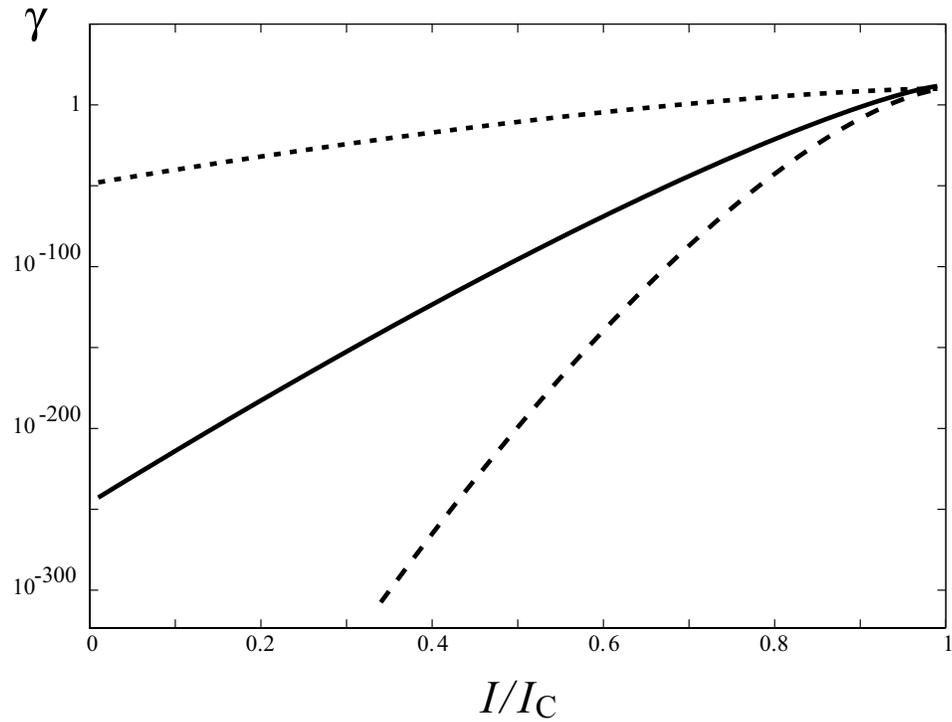}
\caption{\label{fig:gamma} Switching rate $\gamma$ as a function of the relative bias current $I/I_{C}$. The solid line represents the switching rate $\gamma_{\rm{MQT}}$ in the macroscopic quantum tunnelling, while the dotted and the dashed line represents the switching rate $\gamma_{\rm{TA}}$ under the thermal activation model in the temperatures 1~K and 0.1~K, respectively.  The JJ parameters are set to have typical values: Josephson energy \mbox{$E_{J}=10^{-21}$~J} and capacitance $C=50$~femtofarads. $\gamma_{\rm{MQT}}$ is calculated on the high $Q$ limit. The relation between the switching rate $\gamma$ and the height of the current pulse $I$ is nearly linear on the logarithmic scale, which supports the use of the approximation $\gamma(I)=e^{aI+b}$.}
\end{center}
\end{figure}

In this section we study the quantum mechanical arguments leading to the model~\eref{eq:yprob1}.
The switching dynamics are described by a Poisson process where the switching rate $\gamma$ is constant over
the current pulse which has duration $\Delta t$. The probability of switching is then obtained as
\begin{equation}
P(Y=1)=1-e^{-\Delta t \gamma}.
\end{equation}
The duration $\Delta t$ of the current pulse is very short and remains fixed in the experiment. Thus, we consider $\Delta t$ as a part of the switching rate and do not write it explicitly in the subsequent equations.
The relation between the switching rate $\gamma$ and the height of the current pulse $I$ can be written in the form
\begin{equation} \label{eq:rate}
\gamma(I)=e^{ax(I)+b},
\end{equation}
where $x(I)$ is a monotonic function of the bias current $I$ and $a$ and $b$ are constants depending on the physical parameters of the JJ circuit. A commonly used approximation is to assume that the activation threshold amplitude is linearly proportional to current $I$. This leads to the switching rate
\begin{equation} \label{eq:linearrate}
\gamma(I)=e^{aI+b}.
\end{equation}
Although not based on microscopic theory,
the switching rate model~\eref{eq:linearrate} describes the results of typical switching measurements quite well, see Fig.~\ref{fig:gamma}.

In order to obtain the functional form of $x(I)$ in the switching rate model~\eref{eq:rate}
we need a more accurate physical model to describe the system under consideration.
Theoretically, the superconducting electrodes of the JJ sample are modeled
by a complex valued quantum mechanical wave function and the
physical state of the JJ is described by the phase difference over
the junction. The dynamics of the phase difference $\phi$ are
analogous to those of a massive particle in a tilted cosine
potential $U$ under gravitational force,
\begin{equation} \label{eq:washboard}
U=-E_{\rm J} \big( \frac{I}{I_C} \phi+\cos \phi  \big),
\end{equation}
where $E_{\rm J}$ is so-called Josephson energy, $I$ is the magnitude of the applied
bias current, and $I_C$ is the critical current for which the potential changes into monotonic
function. The potential is depicted during a pulse in
Fig.~\ref{fig:expsetup}(d).
The applied current changes the shape of the potential landscape of
a JJ circuit in a controlled way. The phase difference $\phi$ initially
set into one of the local minima becomes a metastable state for finite current $I$;
The state may decay by escape of the phase from the local minimum of the potential to the direction of the steepest descent. This escape event is called switching. In the experiment the voltage over the JJ sample is monitored and the observed voltage pulse indicates that switching has
occurred.

Depending on the temperature, the model~\eref{eq:yprob1} is explained by the thermal activation model of classical physics or by quantum mechanical analysis. In an experiment in temperature above 0.2 K, there will be thermal fluctuations in
the bias current $I$ as the JJ is
connected to the measurement wiring. These fluctuations make the
phase particle to absorb energy and eventually escape from the potential
well. One of the central results in statistical physics is that
the probability for system to be in energy state $E$ at temperature
$T$ is proportional to $\exp(-E/k_{\rm B} T )$, where $k_{\rm B} \approx 1.381\cdot 10^{-23}\rm{J/K}$ is the Bolztman constant.
In order to escape from the well the phase particle needs to have more energy than the barrier
height $\Delta U$. Based on these principles the average escape rate for the potential of equation~\eref{eq:washboard}
can be shown to follow the Kramer's formula
\begin{equation}
\label{eq:gammaTA} \gamma_{\rm{TA}}(I)=\frac{\omega_{P}(I)}{2\pi}\exp(-\Delta
U(I)/k_{\rm B} T ),
\end{equation}
where $\Delta U$ is the height of the potential barrier
\begin{equation}
\Delta U(I) = \frac{4\sqrt{2}E_J}{3} \Big( 1- \frac{I}{I_C}\Big)^{3/2},
\end{equation}
and $\omega_P$ is the characteristic oscillation frequency
\begin{equation}
\omega_{P}(I)=\sqrt{\frac{E_{\rm J}}{C}} \frac{\bar{e}}{\hbar}\Big[ 2 \Big( 1-
\frac{I}{I_C}\Big)\Big]^{1/4},
\end{equation}
where $C$ is the capacitance of the JJ, $\bar{e}=1.602\cdot 10^{-19}$ coulombs is the elementary charge and $\hbar \approx 1.055\cdot 10^{-34}\rm{Js}$ is the reduced Planck's constant.
In above equations $E_{J}$ and $C$  are the unknown quantities -- $I_{C}$ is a function of $E_{J}$. It is easy to see that the switching rate~\eref{eq:gammaTA} can be written in the form $\gamma_{\rm{TA}}(I)=e^{ax(I)+b}$.

When a JJ is cooled to tens of millikelvins
one can still observe the phase particle escaping
from the quasistationary potential~\citep{voss,jackel}.
Since the thermal activation, equation~\eref{eq:gammaTA} yields vanishing switching rate for that temperature range,
the microscopic mechanism must be different from the one described
above. This process is called macroscopic quantum tunnelling~\citep{legget} and to explain it a detailed quantum mechanical analysis is required.  The analysis leads to switching rate~\citep{grabert}
\begin{equation}
\gamma_{MQT}(I)=A(I) e^{-B(I)}. \label{eq:MQTrate}
\end{equation}
The prefactor in equation~(\ref{eq:MQTrate}) is
\begin{equation}
A(I)=\frac{\omega_P(I)}{2\pi}\sqrt{\frac{\Delta
U(I)}{\hbar\omega_P(I)}}\chi (\frac{1}{2Q}), \label{eq:MQTA}
\end{equation}
where $\chi (\alpha )=12\sqrt{6\pi}(1+ c\alpha
+\mathcal{O}(\alpha^2)$, with $c \approx 2.86$,  and $Q$ is the quality factor of the oscillator describing the effective environment of the junction. The exponent reads
\begin{equation}
B(I)=\frac{\Delta U(I)}{\hbar\omega_P(I)}s(\frac{1}{2Q}),
\label{eq:MQTB}
\end{equation}
where $s(\alpha)=36/5[1+45 \zeta (3)/\pi^3 \alpha
+\mathcal{O}(\alpha^2)]$ and $\zeta (3) \approx 1.202$ is the value of the Riemann
zeta function at 3. Again, the switching rate~\eref{eq:MQTrate} can be written in the form $\gamma_{\rm{MQT}}(I)=e^{ax(I)+b}$.

\section{D-optimal designs for complementary log-log regression} \label{sec:doptimal}
Assume that a binary random variable $Y$ is distributed as follows
\begin{align} \label{eq:yprob10}
P(Y=1)&=1-e^{-\exp(ax+b)} \nonumber \\
P(Y=0)&=e^{-\exp(ax+b)},
\end{align}
where $a$ and $b$ are unknown parameters and $x$ is a covariate.
The cumulative distribution function
\begin{equation} \label{eq:Gompertz}
F(z)=1-e^{-\exp(z)},
\end{equation}
where $z=ax+b$,  is known in literature by names Gumbel distribution, Gompertz distribution, LogWeibull distribution,
Fisher-Tippett distribution and extreme value distribution. Whereas, the inverse of \eref{eq:Gompertz}
\begin{equation}
F^{-1}(u)=ax+b=\log(-\log(1-u))
\end{equation}
is called complementary log-log link and is one of the standard link functions in generalized linear models.

The problem is to select such
covariate values $x_{1},x_{2},\ldots,x_{n}$  that the parameters $a$ and $b$ can be estimated from
 the observed data $(x_{i},y_{i}), \; i=1,2,\ldots,n$ in optimal manner.
Here we define the optimality as D-optimality, which equals to maximizing the determinant of the information matrix, and estimate
the parameters by maximum likelihood. Starting from the definition \eref{eq:yprob10} we obtain the log-likelihood of the data $(x_{i},y_{i}), \; i=1,2,\ldots,n$
\begin{equation}
L(a,b)=\sum_{i=1}^{n} \left( y_{i}\log(1-e^{-\exp(ax_{i}+b)})-(1-y_{i})e^{ax_{i}+b}  \right).
\end{equation}
The information matrix
\begin{equation}
\mathbf{J}=-\E
\begin{pmatrix}
\displaystyle \frac{\partial^{2}L}{\partial a^{2}}  & \displaystyle \frac{\partial^{2}L}{\partial a \partial b}\\
\displaystyle \frac{\partial^{2}L}{\partial b \partial a} & \displaystyle \frac{\partial^{2}L}{\partial b^{2}}
\end{pmatrix}
\end{equation}
takes the form
\begin{equation}
\mathbf{J}=
\begin{pmatrix}
\displaystyle \sum_{i=1}^{n}g(z_{i})x_{i}^{2} & \displaystyle \sum_{i=1}^{n}g(z_{i})x_{i}\\
\displaystyle \sum_{i=1}^{n}g(z_{i})x_{i} & \displaystyle \sum_{i=1}^{n}g(z_{i})
\end{pmatrix},
\end{equation}
where $z_{i}=ax_{i}+b$ and
\begin{equation}
g(z)=\frac{e^{2z}}{e^{\exp(z)}-1}.
\end{equation}
The D-optimality is equivalent to finding such $x_{1},x_{2},\ldots,x_{n}$ that
\begin{align} \label{eq:detJ}
\det(\mathbf{J})&=\left[\sum_{i=1}^{n}g(z_{i})x_{i}^{2}\right]\left[\sum_{i=1}^{n}g(z_{i})\right]-\left[\sum_{i=1}^{n}g(z_{i})x_{i}\right]^{2} \nonumber \\
&=\sum_{j=i+1}^{n}\sum_{i=1}^{n-1} g(z_{i})g(z_{j})(x_{i}-x_{j})^{2}
\end{align}
is maximized. It is assumed in the maximization that the true values of $a$ and $b$ are known (see \citep{Minkin:optimal} and \citep{Sitter:robustdesign} for a discussion).

\begin{figure}
\begin{center}
 \includegraphics[width=0.7\columnwidth]{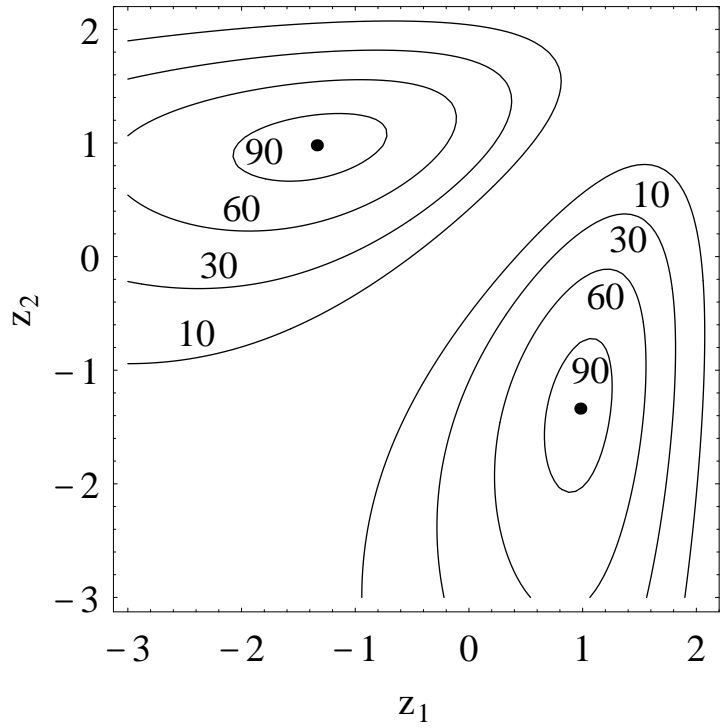}
\caption{Contours of the ratio $100 \cdot \det(\mathbf{J}(z_{1},z_{2}))/\det(\mathbf{J}(z_{1}^{*},z_{2}^{*}))$, where $\det(\mathbf{J}(z_{1},z_{2}))$ is the determinant of the information matrix for the two-point design $(z_{1},z_{2})$. The maximum of $\det(\mathbf{J})$ is achieved at the point $(z_{1}^{*} \approx 0.98,z_{2}^{*} \approx -1.34)$ and at its mirror point.}\label{fig:zcontour}
\end{center}
\end{figure}

First we will find a locally optimal two-point design $n=2$. This simplifies equation~\eref{eq:detJ} to
\begin{equation} \label{eq:detJn2}
\det(\mathbf{J})=\frac{1}{a^{2}}g(z_{1})g(z_{2})(z_{1}-z_{2})^{2}.
\end{equation}
The contour plot of $\det(\mathbf{J})$ as a function $z_{1}$ and $z_{2}$ is depicted in
Fig.~\ref{fig:zcontour}.
The values $z_{1}$ and $z_{2}$ that maximize $\det(\mathbf{J})$
\begin{align} \label{eq:twopointmaximum}
z_{1}^{*} &\approx 0.979633 \nonumber \\
z_{2}^{*} &\approx -1.33774 \nonumber \\
\end{align}
can be solved numerically from \eref{eq:detJn2}. The corresponding response probabilities are
\begin{align} \label{eq:twopointprobs}
F(z_{1}^{*}) &\approx 0.930295 \nonumber \\
F(z_{2}^{*}) &\approx 0.230826,
\end{align}
indicating that the design is asymmetric with respect to the middle point of the response curve.
Since the function is symmetric with respect to exchanging of the variables, the mirror points provide also the
same maximum value.

In order to show that the two-point maximum \eref{eq:twopointmaximum} is a local maximum also when $n>2$, we define functions
\begin{align}
v(z_{1},z_{1})&=g(z_{1})g(z_{2})(z_{1}-z_{2})^{2}\\
v^{'}(z_{1},z_{2})&=\frac{\partial}{\partial z_{1}} v(z_{1},z_{2}).
\end{align}
By inspecting the functional form of $v(z_{i},z_{j})$ we observe that
\begin{align} \label{eq:vmaximum}
v(z_{1}^{*},z_{2}^{*})&=v(z_{2}^{*},z_{1}^{*})&\Rightarrow \textrm{local maximum of }v\\
v(z,z)&=0&\Rightarrow \textrm{local minimum of }v.
\end{align}
We are searching for a local maximum of
\begin{equation}
\det(\mathbf{J})=\frac{1}{a^{2}}\sum_{j=i+1}^{n}\sum_{i=1}^{n-1} v(z_{i},z_{j}).
\end{equation}
For a local extreme of $\det(\mathbf{J})$ it holds
\begin{equation} \label{eq:gradient}
\sum_{j \neq i} v^{'}(z_{i},z_{j})=0 \quad \textrm{for all }i.
\end{equation}
Because $v^{'}(z_{1}^{*},z_{2}^{*})=0$, a solution for equation~\eref{eq:gradient} is obtained if
\begin{equation}
z_{i} \in \{z_{1}^{*},z_{2}^{*}\}  \quad \textrm{for all }i \end{equation}
From equation~\eref{eq:vmaximum} we conclude that a local maximum is obtained, for instance, when
\begin{align}
z_{i}&=z_{1}^{*} \quad \textrm{if $i$ is odd} \nonumber \\
z_{i}&=z_{2}^{*} \quad \textrm{if $i$ is even}.
\end{align}
For the complete proof that the solution~\eref{eq:twopointmaximum} is the global maximum, we refer to the general results by \citet{Ford:canonicaloptimal}. Using a theorem by \citet{Kiefer:equivalence}, \citet{Ford:canonicaloptimal} show for a class of response curves that the D-optimal design is a two-point design. The proof for the complementary log-log regression is obtained as a special case of the general results.

\section{Sequential designs for switching measurements} \label{sec:sequential}
Our goal is to design a fast automated estimation procedure for switching measurements. A fast estimation procedure allows, for example, the tracking of changes in the parameters as a function of changes in environmental variables. The speed of the procedure is measured in terms of the total time used for measuring and estimation. Sequential designs are a natural choice because it is easy to control the current during the experiment. In sequential designs (also known as two-stage and multi-stage designs), we start with initial estimates, collect additional data, obtain new estimates and use them as the initial estimates for the next stage.

In switching measurements, the number of stages can be high: this is a significant difference compared to medical dose response experiments where practical reasons usually limit the number of stages. However, it is not necessarily advisable to choose the number of stages as high as possible, i.e., update the estimates after every two observations.  This is because adding two new observations would have a very small effect on the estimates (especially if the number of observations is already large) but the estimation itself would spend time that could be used for measuring. Additionally, whenever the height of the current pulse is changed, the reconfiguration of the signal generator also takes a time that is considerably longer than the time needed for measuring the response for a single pulse. The optimal updating interval depends also on the computational resources and facilities, and our experiences suggest that it is always reasonable to measure at least 20 times using the same pulse height. Our recommendation is only the lower limit and based on empirically experience on the specific measurement instruments we are using. In the experiment presented  in Section~\ref{sec:real}, we measured 50 times using the same pulse height at the beginning of the experiment and increased this number by 10 \% at every stage so that at the stage 50 we measured 5379 times using the same pulse height.

Obtaining good initial estimates for the parameters is a challenging problem for switching measurements. Naturally, the results from the previous experiments can be used to obtain initial estimates for the parameters but there is no guarantee that this would lead to good initial estimates. The problem of poor initial estimates in the case of logistic regression is considered e.g. by
\citet{King:minimaxdoptimal} and \citet{Sitter:robustdesign}. The authors
propose sophisticated minimax approaches that have good optimality
properties even if the initial estimates are poor. These approaches are
probably applicable also to our problem but because of their
computational complexity, we choose a simpler approach that suits especially for switching measurements. It is usually possible to specify an interval from where the covariates should be taken but the problem is that this interval might be very wide compared to the interval where the probability of response 1 increases, for instance, from 0.01 to 0.99. This implies that initial covariates chosen randomly from the interval may be very inefficient. In the worst case, poorly chosen initial covariates may lead to the situation where maximum likelihood estimates (MLEs) do not exist for the initial data. The existence of MLEs for the model~\eref{eq:yprob10} requires that \citep{Albert:mlexistence}
\begin{align} \label{eq:mlexistence}
\max(x_{i}|y_{i}&=0)>\min(x_{i}|y_{i}=1) \textrm{ and } \nonumber \\ \max(x_{i}|y_{i}&=1)>\min(x_{i}|y_{i}=0).
\end{align}
This condition is fulfilled, for instance, if we have two points where both 0's and 1's are measured as responses. Our approach for the initial estimation aims to finding two such points as quickly as possible. This heuristic procedure is can be presented as follows:
\begin{enumerate}[1.]
\item Use the previous knowledge to construct such an interval
$[x_{\textrm{min}},x_{\textrm{max}}]$ that we can be sure that
$F(x_{\textrm{max}};a,b)-F(x_{\textrm{min}};a,b)$ is close to 1.
Constructing this kind of interval is usually possible even if we have
very little knowledge on the parameters.
\item Use binary search to find such a point $x$ from the interval
$[x_{\textrm{min}},x_{\textrm{max}}]$ that both 0's and 1's are measured as responses. In binary search, we measure the responses at the middle point of the current interval. If only 0's are measured, the middle point is taken as the new starting point of the interval. If only 1's are measured, the middle point is taken as the new end point of the interval. Let $x$ be found as the
middle point of the inteval $[x_{l},x_{u}]$, i.e. $x=(x_{l}+x_{u})/2$.
\item Measure at the point $x+\epsilon$, where $|\epsilon|=(x_{u}-x_{l})/4$. The sign of $\epsilon$ is determined according to the measured response: $\textrm{sign}(\epsilon)=\textrm{sign}(0.5-\bar{y})$, where $\bar{y}$ is the average response for $x$. If $0.5-\bar{y}=0$ the sign of $\epsilon$ is chosen randomly.
\item If MLEs exist for the data measured so far, proceed to the maximum
likelihood estimation. Otherwise, divide $\epsilon$ by two and return to Step 3.
\end{enumerate}
Step 2 requires that we measure at least two times at each point. In practice, it is preferable to measure at least tens of times at each point because of time needed for the reconfiguration of the signal generator. When the binary search in Step 2 converges we have found one of the two points required for the maximum likelihood estimation. The other point that is needed is then found in the neighbourhood of point $x$ using again binary search.  Due to use of binary search in Steps 2 and 3, the procedure works reliably and efficiently.  At every step of binary search, the length of the current interval is divided by two. Consequently, the procedure converges exponentially fast regardless of the choice of the initial interval.

After the heuristic procedure has converged the experiment continues sequentially: The current values of $\hat{a}$ and $\hat{b}$ are used to determine the optimal covariates for the next stage, measurements are performed, and finally maximum likelihood estimation is applied to update values of $\hat{a}$ and $\hat{b}$. In switching measurements, one is also interested in
parameters that directly describe the location and the scale of the
response curve in addition to parameters $a$ and $b$.
Let us define the middle point $\theta$ and width of the curve
$\lambda$, as follows
\begin{align}
\theta&=\frac{1}{a}(F^{-1}(0.5)-b)\\
\lambda&=\frac{1}{a}(F^{-1}(0.9)-F^{-1}(0.1)).
\end{align}
These two parameters are frequently used in switching measurements to characterize the response curve. The covariance matrix of $(\theta \; \lambda)^{\intercal}$ is given by
\begin{equation}
\mathbf{A}\mathbf{J}^{-1}\mathbf{A}^{\intercal},
\end{equation}
where
\begin{equation}
\mathbf{A}=
\begin{pmatrix}
\displaystyle  \frac{\partial \theta}{\partial a} & \displaystyle  \frac{\partial \theta}{\partial b} \vspace{0.1cm}\\
\displaystyle  \frac{\partial \lambda}{\partial a} & \displaystyle  \frac{\partial \lambda}{\partial b}
\end{pmatrix}
=
\begin{pmatrix}
\displaystyle  \frac{b-F^{-1}(0.5)}{a^{2}} & \displaystyle  -\frac{1}{a}\\
\displaystyle  -\frac{F^{-1}(0.9)-F^{-1}(0.1)}{a^{2}} & \displaystyle  0
\end{pmatrix}.
\end{equation}

It is often practical to define the stopping rule for the data acquisition using the variances of $\theta$ and $\lambda$. Special attention should be paid to the stability of the measurement environment. Instability, such as a change in the temperature during the experiment, causes an increase in the variances. This emphasizes the need for fast and reliable experimental design.

\section{Examples with real and simulated data } \label{sec:real}
We measured a sample consisting of an aluminium--aluminium oxide--aluminium JJ circuit in a dilution refrigerator at 20~mK temperature.
The sample parameters $a$ and $b$ depend on the area of the JJ and the thickness of the oxide layer. The sample fabricated
 involves a JJ whose area is 1 times 2 micrometers and the oxide layer is ten nanometres thick.
The sample was connected to computer controlled measurement electronics in order to apply the current pulses and record the resulting voltage pulses.
The resistance of the sample at room temperature suggests that a pulse of 300~nA always causes a switching (response~1). This gives the upper limit for the
initial estimation. The lower limit for the initial estimation, 200~nA,  can be roughly estimated from the dimensions of the JJ.
The sequential design presented in Section~\ref{sec:sequential} was used to construct the applied current pulse sequences in real time. Each stage consisted of measuring phase and estimation phase. Since the reconfiguration of our signal generator takes a relatively long time compared to the length of a single pulse, we used sequences of tens or hundreds of pulses to collect data at each stage. In the setup the pulse lengths are on the order of milliseconds. In order to keep the time needed for computations much below the actual measurement time, the length of the pulse sequence should grow exponentially. We started with a sequence of 50 pulses per point and let the number of pulses grow 10 percent at each stage. In the test experiments, the practically sufficient accuracy of the estimates was reached after 20--30 stages, implying that 6000--17000 pulses were needed in total. In our setup, the experiment (50~stages) took 8 minutes 16 seconds from which 1~minute 30~seconds were used for estimation.

\begin{figure}
\begin{center}
\includegraphics[width=0.98\columnwidth]{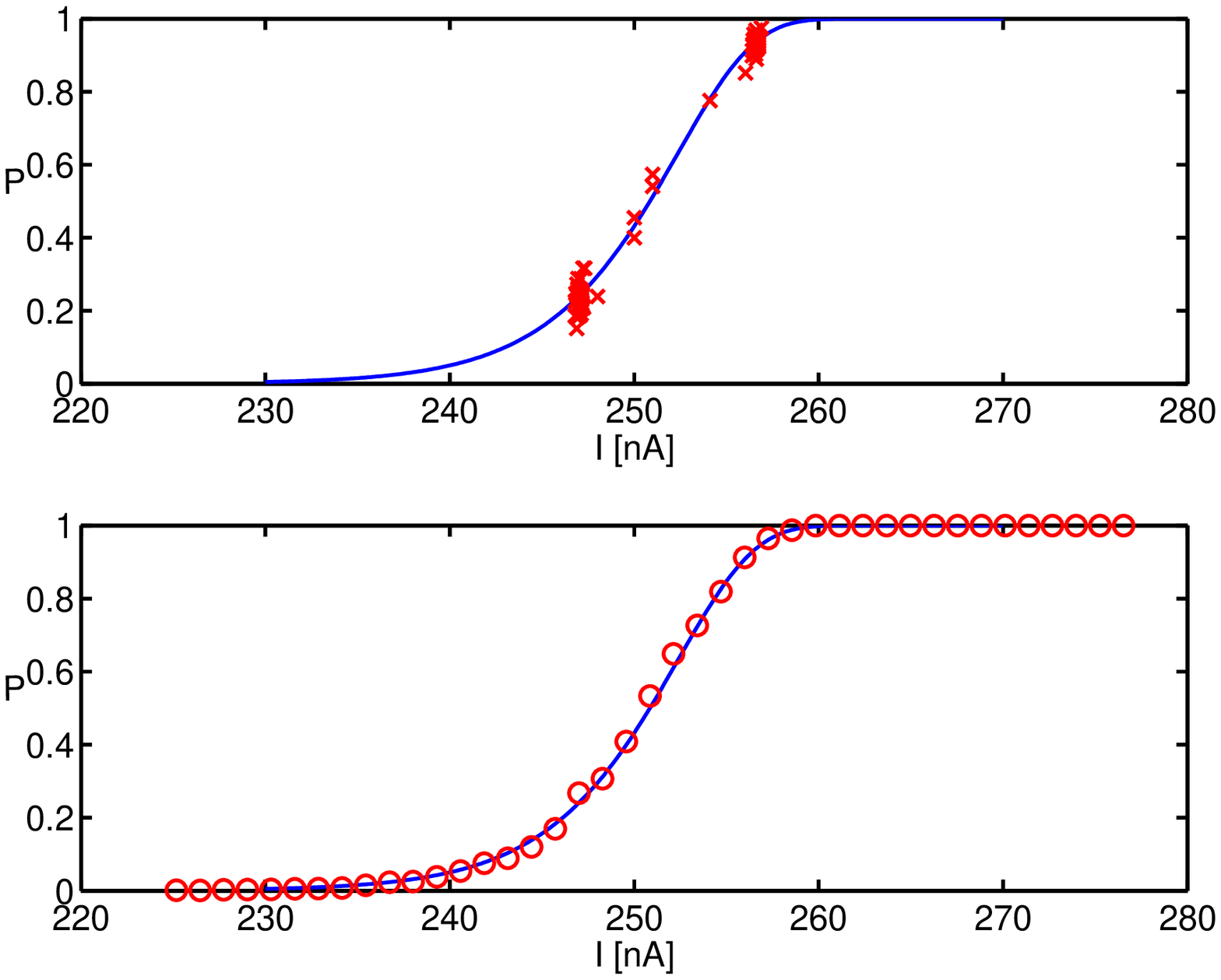}
\caption{Estimated response curve for the measured JJ. Upper panel: estimated response curve with data used in the estimation (pulse heights, $I$, and corresponding empirical probabilities, $P$). Lower panel: estimated response curve and verification points obtained as an average of responses of 2000 independent pulses.}\label{fig:expdata}
\end{center}
\end{figure}

Table~\ref{tab:expestim} and Fig.~\ref{fig:expdata} illustrate the estimation of the unknown parameters in the experiment. In the lower panel of Fig.~\ref{fig:expdata} the functional form of the response curve is verified empirically. The response curve is compared with individual response probabilities that are independently estimated using 2000 pulses per point. The results confirm that the response curve is given by the Gumbel distribution as suggested by the theory.

\begin{table}
\caption{Estimates and their standard deviations for the measured JJ.  \label{tab:expestim}}
\begin{tabular}{rrrrrr}
Stage &  n & \multicolumn{1}{c}{$\hat{a}$} & \multicolumn{1}{c}{std($\hat{a}$)} & \multicolumn{1}{c}{$\hat{b}$} & \multicolumn{1}{c}{std($\hat{b}$)}\\ \hline
1 &100 &    \multicolumn{2}{c}{No MLE}  & \multicolumn{2}{c}{No MLE}  \\
2 &210 &    \multicolumn{2}{c}{No MLE} & \multicolumn{2}{c}{No MLE} \\
3 &332 &    0.3800 &0.1933 &-95.5966 &48.4418 \\
4 &466 &    0.2585 &0.0392 &-65.1942 &9.8369 \\
5 &614 &    0.2301 &0.0224 &-58.1248 &5.6501 \\
6 &776 &    0.2520 &0.0177 &-63.6371 &4.4625 \\
7 &954 &    0.2474 &0.0148 &-62.5090 &3.7268 \\
8 &1150 &   0.2368 &0.0126 &-59.8039 &3.1855 \\
9 &1366 &   0.2366 &0.0112 &-59.7514 &2.8222 \\
10 &1604 &  0.2426 &0.0102 &-61.2916 &2.5744 \\
15 &3200 &  0.2350 &0.0067 &-59.3476 &1.7022 \\
20 &5762 &  0.2427 &0.0050 &-61.2986 &1.2637 \\
25 &9890 &  0.2433 &0.0038 &-61.4761 &0.9572 \\
30 &16550 & 0.2424 &0.0029 &-61.2533 &0.7359 \\
40 &44578 & 0.2410 &0.0017 &-60.8778 &0.4430 \\
50 &117288 &0.2402 &0.0011 &-60.6282 &0.2708 \\
\\
Stage &  n & \multicolumn{1}{c}{$\hat{\theta}$} & \multicolumn{1}{c}{std($\hat{\theta}$)} & \multicolumn{1}{c}{$\hat{\lambda}$} & \multicolumn{1}{c}{std($\hat{\lambda}$)} \\ \hline
1 &100 &    \multicolumn{2}{c}{No MLE}  & \multicolumn{2}{c}{No MLE}  \\
2 &210 &    \multicolumn{2}{c}{No MLE} & \multicolumn{2}{c}{No MLE} \\
3 &332 &    250.5738 &0.2519 &8.1158 &4.1282 \\
4 &466 &    250.8324 &0.3037 &11.9342 &1.8080 \\
5 &614 &    251.0506 &0.2984 &13.4066 &1.3070 \\
6 &776 &    251.0632 &0.2519 &12.2392 &0.8591 \\
7 &954 &    251.1471 &0.2327 &12.4655 &0.7432 \\
8 &1150 &   251.0128 &0.2219 &13.0259 &0.6933 \\
9 &1366 &   250.9891 &0.2058 &13.0362 &0.6148 \\
10 &1604 &  251.1269 &0.1874 &12.7136 &0.5330 \\
15 &3200 &  250.9385 &0.1386 &13.1228 &0.3751 \\
20 &5762 &  251.1056 &0.1019 &12.7110 &0.2610 \\
25 &9890 &  251.1342 &0.0780 &12.6756 &0.1965 \\
30 &16550 & 251.1604 &0.0605 &12.7233 &0.1522 \\
40 &44578 & 251.1016 &0.0372 &12.7992 &0.0927 \\
50 &117288 &250.8531 &0.0231 &12.8395 &0.0571 \\
\end{tabular}
\end{table}

The performance of the sequential design is also studied in a simulation example. The simulation aims to imitate the real data example .  The response variable $Y$ is generated from the model~\eref{eq:yprob10} with parameters $a=0.24$ and \mbox{$b=-61$} which are motivated by the actual experiment presented above. Our goal is to estimate these parameters when we initially know only that at point $x_{\textrm{min}}=200$ the response is 0 with high probability and at point $x_{\textrm{max}}=300$ the response is 1 with high probability. The estimation starts using the heuristic procedure described in Section~\ref{sec:sequential} and continues as maximum likelihood estimation. In the initial estimation, we measure 25 times at each point. In the sequential maximum likelihood estimation,
the parameter estimates are updated when the total number of observations has increased by 20~\% from the previous update. The simulation is repeated 500 times and maximum likelihood estimates $\hat{a}$ and $\hat{b}$ are recorded for different numbers of observations $n=50, n=100,\ldots,n=20000$ . The sequential design is compared to the theoretically D-optimal design where the optimal covariate values $x_{1}^{*}=(z_{1}^{*}+61)/0.24=258.25$ and $x_{2}^{*}=(z_{2}^{*}+61)/0.24=248.59$ are assumed to be known already at the beginning of the experiment. As a performance measure we calculate the mean squared errors (MSE) between the estimates and the true parameter values.

Table~\ref{tab:simuestim} summarizes the obtained MLEs and their standard deviations. The reported numbers are means from 500 simulations. The MSEs between the estimates and the true parameter values are reported in Table~\ref{tab:simumse}. The MSEs are calculated only from those simulation runs where MLEs exist. Naturally, the theoretically optimal design results smaller MSE values than the sequential design but it can be also seen that for large values of $n$ the performance of the sequential design approaches to the performance of the theoretically optimal design. The results demonstrate the practical importance of good experimental design and the significance of reliable estimation of initial parameters. We conclude that the proposed sequential design performed reasonably well compared to the theoretically D-optimal design when $n>200$. In switching  measurements, the number of observations is always in thousands. Comparison of estimates and their standard deviations in Table~\ref{tab:expestim} and Table~\ref{tab:simuestim} does not reveal any major differences between the experimental data and simulated data. This supports the practical usability of the sequential design.

\begin{table}
\caption{Means of estimates and their standard deviations from 500 simulation runs.  \label{tab:simuestim}}
\begin{tabular*}{\columnwidth}{rrrrrrrrr}
& \multicolumn{2}{c}{Sequential} & \multicolumn{2}{c}{Optimal} & \multicolumn{2}{c}{Sequential} & \multicolumn{2}{c}{Optimal}\\
n & \multicolumn{1}{c}{$\hat{a}$} & \multicolumn{1}{c}{std($\hat{a}$)} & \multicolumn{1}{c}{$\hat{a}$} & \multicolumn{1}{c}{std($\hat{a}$)} & \multicolumn{1}{c}{$\hat{b}$} & \multicolumn{1}{c}{std($\hat{b}$)} & \multicolumn{1}{c}{$\hat{b}$} & \multicolumn{1}{c}{std($\hat{b}$)}\\ \hline
50 &0.4342 &0.3953 &0.2454 &0.0659 &-109.6947 &99.0491 &-62.4331 &16.8447 \\
100 &0.2786 &0.1151 &0.2476 &0.0387 &-70.7275 &28.8748 &-62.9339 &9.8681 \\
200 &0.2482 &0.0373 &0.2420 &0.0262 &-63.0732 &9.4241 &-61.4907 &6.6879 \\
300 &0.2448 &0.0254 &0.2418 &0.0213 &-62.2302 &6.4475 &-61.4518 &5.4436 \\
400 &0.2433 &0.0203 &0.2413 &0.0184 &-61.8373 &5.1676 &-61.3297 &4.6997 \\
500 &0.2425 &0.0176 &0.2405 &0.0164 &-61.6395 &4.4892 &-61.1274 &4.1928 \\
700 &0.2415 &0.0145 &0.2408 &0.0139 &-61.3900 &3.6946 &-61.2015 &3.5441 \\
1000 &0.2418 &0.0120 &0.2409 &0.0116 &-61.4646 &3.0482 &-61.2335 &2.9653 \\
1500 &0.2410 &0.0096 &0.2406 &0.0095 &-61.2491 &2.4610 &-61.1591 &2.4188 \\
2000 &0.2411 &0.0083 &0.2408 &0.0082 &-61.2892 &2.1225 &-61.2076 &2.0953 \\
3000 &0.2408 &0.0068 &0.2405 &0.0067 &-61.2001 &1.7240 &-61.1375 &1.7089 \\
4000 &0.2406 &0.0058 &0.2405 &0.0058 &-61.1488 &1.4894 &-61.1361 &1.4799 \\
5000 &0.2405 &0.0052 &0.2402 &0.0052 &-61.1338 &1.3305 &-61.0467 &1.3223 \\
7000 &0.2404 &0.0044 &0.2401 &0.0044 &-61.0939 &1.1225 &-61.0178 &1.1170 \\
10000 &0.2402 &0.0037 &0.2401 &0.0037 &-61.0621 &0.9378 &-61.0184 &0.9344 \\
15000 &0.2402 &0.0030 &0.2400 &0.0030 &-61.0505 &0.7648 &-61.0077 &0.7629 \\
20000 &0.2401 &0.0026 &0.2401 &0.0026 &-61.0296 &0.6619 &-61.0128 &0.6607 \\\\
& \multicolumn{2}{c}{Sequential} & \multicolumn{2}{c}{Optimal} & \multicolumn{2}{c}{Sequential} & \multicolumn{2}{c}{Optimal}\\
n& \multicolumn{1}{c}{$\hat{\theta}$} & \multicolumn{1}{c}{std($\hat{\theta}$)} & \multicolumn{1}{c}{$\hat{\theta}$} & \multicolumn{1}{c}{std($\hat{\theta}$)}& \multicolumn{1}{c}{$\hat{\lambda}$} & \multicolumn{1}{c}{std($\hat{\lambda}$)} &\multicolumn{1}{c}{$\hat{\lambda}$} & \multicolumn{1}{c}{std($\hat{\lambda}$)} \\ \hline
50 &254.3894 &7.2977 &252.7209 &1.3866 &21.0825 &38.7785 &13.1512 &3.6519 \\
100 &253.1132 &1.7524 &252.6023 &0.8089 &14.3413 &7.8859 &12.7611 &2.0261 \\
200 &252.7223 &0.5869 &252.5675 &0.5638 &12.7674 &1.9482 &12.8914 &1.4084 \\
300 &252.6707 &0.4449 &252.5894 &0.4589 &12.7352 &1.3232 &12.8599 &1.1416 \\
400 &252.6603 &0.3783 &252.5911 &0.3972 &12.7684 &1.0674 &12.8631 &0.9864 \\
500 &252.6529 &0.3393 &252.5970 &0.3554 &12.7853 &0.9307 &12.8892 &0.8837 \\
700 &252.6479 &0.2892 &252.6178 &0.2994 &12.8159 &0.7700 &12.8514 &0.7427 \\
1000 &252.6554 &0.2433 &252.6364 &0.2501 &12.7847 &0.6323 &12.8315 &0.6196 \\
1500 &252.6535 &0.2003 &252.6472 &0.2041 &12.8202 &0.5134 &12.8361 &0.5058 \\
2000 &252.6544 &0.1741 &252.6507 &0.1766 &12.8081 &0.4419 &12.8224 &0.4373 \\
3000 &252.6413 &0.1428 &252.6444 &0.1443 &12.8203 &0.3597 &12.8329 &0.3573 \\
4000 &252.6417 &0.1240 &252.6472 &0.1249 &12.8284 &0.3111 &12.8305 &0.3093 \\
5000 &252.6426 &0.1110 &252.6423 &0.1118 &12.8301 &0.2780 &12.8474 &0.2771 \\
7000 &252.6454 &0.0940 &252.6379 &0.0945 &12.8368 &0.2348 &12.8517 &0.2342 \\
10000 &252.6435 &0.0788 &252.6366 &0.0791 &12.8420 &0.1963 &12.8504 &0.1959 \\
15000 &252.6401 &0.0644 &252.6376 &0.0646 &12.8432 &0.1601 &12.8518 &0.1600 \\
20000 &252.6391 &0.0558 &252.6391 &0.0559 &12.8470 &0.1387 &12.8503 &0.1385 \\
\end{tabular*}
\end{table}

\begin{table}
\caption{Mean squared errors (MSE) for parameter estimates $\hat{a}$ and $\hat{b}$ as functions of number of observations $n$. The proposed sequential design and the theoretically D-optimal design are compared. Reported MSEs are means from 500 simulation runs. \label{tab:simumse}}
\begin{tabular}{rrrrr}
 & \multicolumn{2}{c}{MSE of $\hat{a}$} & \multicolumn{2}{c}{MSE of $\hat{b}$} \\
n &  \multicolumn{1}{c}{Sequential} & \multicolumn{1}{c}{Optimal} & \multicolumn{1}{c}{Sequential} &\multicolumn{1}{c}{Optimal} \\ \hline
50 &0.206843 &0.002715 &12983.570197 &179.184180 \\
100 &0.021449 &0.001546 &1348.770300 &101.345741 \\
200 &0.001771 &0.000667 &112.423833 &43.457153 \\
300 &0.000686 &0.000477 &44.089057 &31.041668 \\
400 &0.000439 &0.000373 &28.286722 &24.314540 \\
500 &0.000316 &0.000297 &20.389242 &19.364127 \\
700 &0.000211 &0.000196 &13.672440 &12.766761 \\
1000 &0.000139 &0.000132 &9.035246 &8.643545 \\
1500 &0.000093 &0.000080 &6.046665 &5.201665 \\
2000 &0.000077 &0.000063 &4.993876 &4.132176 \\
3000 &0.000049 &0.000045 &3.198680 &2.937851 \\
4000 &0.000037 &0.000032 &2.390940 &2.100386 \\
5000 &0.000030 &0.000024 &1.946899 &1.584166 \\
7000 &0.000021 &0.000017 &1.379387 &1.116941 \\
10000 &0.000015 &0.000012 &0.961786 &0.811790 \\
15000 &0.000010 &0.000008 &0.634116 &0.523743 \\
20000 &0.000007 &0.000006 &0.473926 &0.399793 \\
\end{tabular}
\end{table}

\section{Conclusion} \label{sec:conclusion}
We have considered optimal designs for switching measurements of superconducting Josephson junction circuits. The D-optimality criterion suggests that the measurements at points where $ax_{1}+b=0.979633$ and $ax_{2}+b=-1.33774$ will yield a maximal information about the estimated parameters assuming that the probability distribution of the measurement values are given by Eq.~\eref{eq:yprob1}. As a practical method for switching measurements, we propose a sequential design that combines heuristic search for initial estimates and maximum likelihood estimation. The fast data acquisition allows tracking of fast environmental changes. The presented experimental design is now in daily use in Low Temperature Laboratory at Helsinki University of Technology.

The current work opens several directions for the future research. Besides D-optimality we may also consider other optimality criteria. These criteria likely lead to designs different from the D-optimal design but because of the complicated functional forms it may be difficult to derive analytical results.

The problem of choosing the number of measurements per stage was approached empirically in this paper. A more systematic approach would determine the optimal number of measurements per stage as a function of the observed data. It seems reasonable to assume that each measurement takes a certain time and the change of the measurement point takes additional fixed time. The problem is to choose the number of measurements per stage in such a way that the expected total time needed to achieve the required accuracy is minimized. This is a non-trivial problem and probably only approximate solutions are reachable.

The time used for the experiment could be further reduced if the estimation and the measuring were done simultaneously. This requires good coordination of data processing and knowledge on the time needed for each operation. In sequential designs, the simultaneous estimation and measuring implies that the current estimates are not based on all data measured so far but all data processed so far.

Although the presented complementary log-log regression model describes well some JJ circuits involving several junctions,
we probably need more flexible models  for complicated structures of JJ. The modelling can be approached from two different directions: we may derive the exact physical model and try to find the optimal design for it, or we may start with a flexible parametric model for which the optimal design is known and estimate the model parameters from the measured data.

Statistical methods can be also applied to the modelling of JJ parameters as a function of environmental variables, such as temperature. This direction of research leads to adaptive sequential designs where the challenge is not only to estimate the parameters but also to track their changes in optimal way.

%Overall, we see switching measurements as a novel and inspiring application field %of statistics.

\bibliographystyle{chicago_mod}
\bibliography{optdesign}

\end{document}